\documentstyle[epsf,12pt]{ioplppt}
\begin{document}
\input psfig
\nopagebreak
\vspace{-1.1in}
\begin{flushright}
UR-1460, ER-40685-889  \\
UK PHENOMENOLOGY WORKSHOP\\
A Workshop on HERA Physics\\
17th to 23rd September, 1995
\end{flushright}
\vspace{-0.5in}

\title{A Measurement of $R ={\sigma_L}/{\sigma_T}$ in Deep Inelastic
Neutrino-Nucleon Scattering at the Tevatron}

\author{U~K~Yang,~P~Auchincloss,~P~de~Barbaro,~A~Bodek,~H~Budd,~D~A~Harris,\\
 W~K~Sakumoto\dag;~W~C~Lefmann,~R~Steiner\ddag;~M~J~Oreglia,~B~A~Schumm\S; \\
     R~A~Johnson,~M~Nussbaum,~L~Perera,~M~Vakili$\Vert$;~C~G~Arroyo,\\
A~O~Bazarko,~J~Conrad,~J~H~Kim,~B~J~King,~S~Koutsoliotas,~C~McNulty,\\
S~R~Mishra,~Z~Quintas,~A~Romosan,~F~J~Sciulli,~W~G~Seligman,\\
M~H~Shaevitz,~~Spentzouris,~E~G~Stern\P;~R~H~Bernstein,~G~Koizumi,\\
M~J~Lamm,~~Marsh,~K~S~McFarland$^+$;~T~Bolton,~W~B~Lowery,~D~Naples$^*$;\\
    R~B~Drucker$\sharp$;~T~Kinnel,~W~H~Smith$\natural$;~and~P~Nienaber$^\#$ }

\address{ ~{\bf {(Presented by U~K~Yang for the CCFR/NuTeV Collaboration)}}\\
\dag ~University of Rochester, Rochester, NY 14627 USA \\
\ddag ~Adelphi University, Garden City, NY 11530 USA \\
\S ~University of Chicago, Chicago, IL 60637 USA \\
$\Vert$ University of Cincinnati, Cincinnati, OH 45221 USA \\
\P~Columbia University, New York, NY 10027 USA \\	
$^+$Fermi National Accelerator Laboratory, Batavia, IL 60510 USA \\
$*$~Kansas State University, Manhattan, KS 66506 USA \\
$\sharp$~University of Oregon, Eugene, OR 97403 USA \\
$\natural$~University of Wisconsin, Madison, WI 53706 USA \\
$^\#$~Xavier University, Cincinnati, OH 45207 USA }

\vspace{-0.07in}
\begin{abstract} 
Measurements of neutrino-nucleon and antineutrino-nucleon
differential cross sections using the CCFR neutrino detector
at Fermilab have been used to extract preliminary
values of $R ={\sigma_L}/{\sigma_T}$
in the kinematic region $0.01<x<0.6$, and $4<Q^2<300$ GeV$^2$.
The new data provide the first measurements of $R$ in the $x<0.1$
region. The $x$ and $Q^2$ dependence of $R$ is compared with a QCD based fit
to previous data. The QCD fit, which provides
an estimate of $R$ in small $x$ region where $R$ has
not been previously measured, is in good agreement with the new CCFR data.\\
{\centering {\it (\underline {To be published in Jou. Phys. G})}}
\end{abstract}
\vspace{-0.25in}

\nopagebreak

\section {Introduction}

    The ratio $R ={\sigma_L}/{\sigma_T}$ of the longitudinal and 
transverse absorption cross sections in deep inelastic lepton-nucleon 
scattering provides information
about the transverse momentum and spin of the nucleon constituents. 
Within the theory of quantum chromodynamics (QCD), the nucleon
constituents are spin 1/2 quarks and spin 1 gluons. 
In leading order QCD, $R = 0$, since the
quarks have no transverse momentum. In the next to leading order
formalism (NLO), to first order in $\alpha_s$, $R$ is non-zero
because of transverse momentum associated with gluon emission~\cite{qcd1}.
The NLO QCD prediction is given 
by an integral over the quark and gluon distribution and
is proportional to $\alpha_s$.
A measurement of $R$ is a test of
perturbative QCD at large $x$, and a clean probe of the gluon density 
at small $x$ where the quark contribution is small. 

Poor knowledge of $R$ , especially at small $x$,
results in uncertainties 
in the structure functions extracted from deep inelastic scattering
cross sections.  The most precise previous
measurements of $R$ are from SLAC electron scattering
experiments~\cite{slacr}, and are at high $x$ and low $Q^2$,
 where non-perturbative contributions
(e.g. target mass and higher twist effect) are important. Therefore,
extraction of structure functions at large $Q^2$ or small $x$ must rely on
good measurements of $R$, or on reasonable estimates of $R$ in the
kinematic region where $R$ has not been measured. 

 We report here on
a new extraction of $R$ from measurements of neutrino and antineutrino
differential cross sections in the CCFR neutrino detector. 
We compare the data (particularly at small $x$)
 to the prediction of a QCD based fit to $R$ by
Bodek, Rock and Yang~\cite{rpaper}.
Extrapolation of such a QCD based fit to smaller values of $x$
can be also used to obtain estimates of $R$ in the HERA region
where $R$ has not been measured.

\section {$R$ calculation within QCD}

In this section we provide a brief description of the QCD based
model of Bodek, Rock and Yang~(BRY). They have used a calculation~\cite{qcd2l}
of the QCD contribution to $R$ to order $\alpha_s^2$ (NNLO). 
The contribution of massive heavy quark
(charm) effects to $R$ has been included within the photon gluon fusion 
process~\cite{qcd2h}. These higher order QCD corrections and heavy
charm quark effects  are important at small $x$.
In addition, the SLAC data on $R$
indicate that there are additional non-perturbative
$1/Q^2$ and $1/Q^4$ contributions 
at low $Q^2$~\cite{slacr,rpaper}. One of these contributions,  target mass 
effects, dominates at low $Q^2$ and large $x$. 
A comparison of the calculation (including both the heavy charm
contribution to order $\alpha_s^2$ and the Georgi-Politzer 
target mass corrections~\cite{gptm})
to the SLAC data indicates that additional 
small higher twist contribution at low $Q^2$ are required~\cite{rpaper}.
A parameterization of the higher twist effects at low $Q^2$
should be constrained such that
$R$ approaches zero at the photoproduction limit ($Q^2 = 0$).
BRY find that 
a simple empirical parameterization of the higher twist
contribution to $R$ (of the form $A[(Q^2-B)/(Q^4+C)]$) 
fits the SLAC data very well. The best BRY fit to the all previous
data~\cite{slacr,otherr} (for $Q^2 > 0.4$ GeV$^2$ ), 
using the GRV94~\cite{grv94}
 parton distribution for the QCD
contribution, yields $A=0.37$, $B=0.82$, 
and $C=0.54$ ($\chi^2/dof$ is 21.7/19). 
Figure~1 shows some of the previous data and the model predictions
calculated using various parton distributions. All curves
include QCD to order $\alpha_s^2$,
the Georgi-Politzer target mass correction and the empirical small higher
twist contribution. The curves labeled with (h) also include
the effects of the charm quark mass. Also shown for comparison
are the extrapolation of a previous empirical fit (Rworld) for $R$ used
by the SLAC, CCFR, and NMC collaborations. The QCD model predictions
with various parton distributions~\cite{grv94,mrs} all
yield similar results for $R$ at large $x$ ($>0.1$). 
At small $x$,  the predictions of the model using
either the MRSD$'_-$ and the MRSA distributions are very different
because of the different gluon distributions. Note that
target mass effects at small $x$ are negligible, and the 
small empirical higher twist contribution is assumed to be 
independent of $x$.  The model calculated with the GRV94
parton distributions (solid curve, figure~1) is probably the best available 
estimate
for extracting $F_2$ in regions where $R$ has not been measured. We proceed
to test the models shown in figure 1 against the new CCFR data 
in the section below.

\begin{figure}
\centerline{\psfig{figure=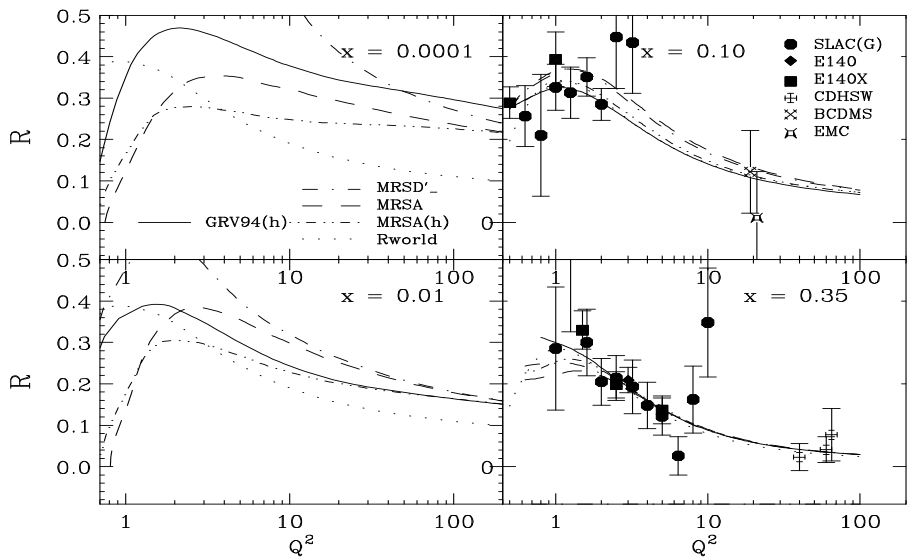,width=4.7in,height=2.88in}}
\caption{$R$ at fixed $x$ vs $Q^2$: A comparison of the BRY prediction of
$R$ calculated using various parton distributions (including
target mass and empirical higher twist) to data. 
A massive heavy (charm) quark contribution is included 
in the MRSA(h) and GRV94(h) prediction. Also shown is Rworld, which
is a previous empirical parameterization of $R$ data in the SLAC region.}
\end{figure}

\section {A Preliminary CCFR measurement of $R$}

The data sample includes two CCFR data runs (E744 and E770) collected using
the Fermilab Tevatron Quad-Triplet neutrino beam. The wide-band beam
is composed of $\nu_\mu$ and $\overline{\nu}_\mu$ with energies up to 600 GeV.
The CCFR neutrino detector~\cite{calib} consists of an
unmagnetized steel-scintillator target calorimeter
instrumented with drift chambers.
The hadron energy resolution is 
$\Delta E/E = 0.85/\sqrt{E}$(GeV). The target is followed by a solid iron
toroidal magnet  muon spectrometer 
which measures muon momentum with a resolution
$\Delta p/p = 0.11$. The relative flux
at different energies is obtained from the events 
with low hadron energy, $\nu < 20$ GeV, and is normalized so that
the neutrino total cross section equals the world average $\sigma^{\nu N}/E=
(0.674\pm0.014)\times10^{-38}$ cm$^2$/GeV, and $\sigma^{\overline{\nu} N}
/\sigma^{\nu N}=0.504\pm0.005$. 

\begin{figure}
\centerline{\psfig{figure=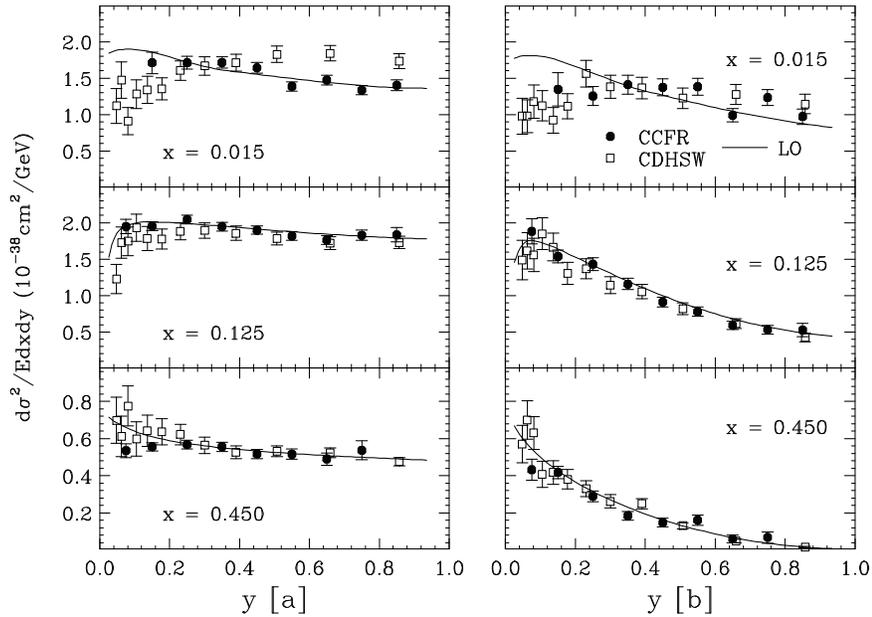,width=4.7in,height=3.2in}}
\caption{Preliminary differential cross section (statistical error only) 
for neutrino[a] and antineutrino[b]
at $E_\nu=150$ GeV: Preliminary CCFR data (solid circles)
are compared to the QCD LO prediction and previous CDHSW data (open squares).}
\end{figure}

Preliminary values of $R$ have been
 extracted from the differential cross sections 
in neutrino charged current scattering. 
The differential cross sections are written as
\begin{equation}
 \frac{d^2\sigma}{Edxdy} =
   \frac{G_F^2M}{\pi} \left[ F_2(x,Q^2) \left( 1-y+ \frac {y^2}{2(1+R)} 
   \right)
\pm xF_3(x,Q^2) \left( y-\frac{y^2}{2} \right) \right]
\end{equation}
where $y=\nu/E_\nu$ is the inelasticity,
and $R={\sigma_L}/{\sigma_T}=F_2(1+Q^2/\nu^2)/2xF_1 -1$.
In the quark-parton model, $2xF_1^{\nu,\overline{\nu}}=q+\overline{q}$
and $xF_3^{\nu,\overline{\nu}}=(q-\overline{q})\pm 2(s-c)$ 
on an isoscalar target.

The data sample consists of
1,280,000 $\nu_{\mu}$ and 270,000 $\overline{\nu}_{\mu}$ events
after fiducial and kinematic cuts ($P_{\mu}>15$ GeV, $\theta_{\mu} <.150$,
$\nu > 10$ GeV, $Q^2 > 1$ GeV$^2$, and $E_{\nu} > 30$ GeV). Dimuon
events are removed because of the ambiguous identification of the leading muon
for high $y$ events. The data are corrected by Monte Carlo 
for cuts,  acceptance and resolution smearing.
The differential cross sections are determined in bins of $x$, $y$ 
and $E_{\nu}$
($0.01 < x < 0.60$, $0.05<y<0.95$, and $30< E_\nu <360$ GeV). Figure~2 shows
the differential cross sections at $E_\nu=150$ GeV. The differential
cross sections agree with a LO QCD model which predicts a 
quadratic $y$ dependence at small $x$ for neutrino and anti-neutrino
and a flat $y$ distribution at high $x$ for the neutrino cross section.
A comparison of CCFR and CDHSW differential cross sections~\cite{otherr} 
indicates good agreement at high $x$, and poor agreement at small $x$, as 
shown in Figure~2.

  The values of $R$ are extracted 
from linear fits to $F$ vs $\epsilon$ at fixed $x$ and $Q^2$ bins
through the following relation,
\begin{equation}
 F(x,Q^2,\epsilon) = \frac {\pi (1-\epsilon) }{y^2 G_F^2 ME}
 \left[ \frac{d^2\sigma ^{\nu}}{dxdy} + \frac{d^2\sigma^{\overline{\nu}}}{dxdy}
 \right] 
 = 2xF_1(x,Q^2) \left[ 1+\epsilon R(x,Q^2) \right]
\end{equation}
which assumes $xF^{\nu}_3=xF^{\overline \nu}_3$. 
Here $\epsilon\simeq2(1-y)/(1+(1-y)^2)$ is the polarization of virtual
 $W$ boson.
Corrections for electroweak radiative effects (Bardin), 
the $W$ boson propagator,
 and isoscaler target (the 6.85\% excess 
of neutron over proton in iron) are applied. The correction for
$\Delta xF_3=xF^{\nu}_3-xF^{\overline \nu}_3$ (which is important at
small $x$) requires a knowledge
of the strange sea and charm production mass effects (e.g. slow rescaling).
The strange sea distribution,
extracted to NLO from CCFR dimuon data~\cite{nlost}
has been used.
The fits for $R$ at each $x$ and $Q^2$ have reasonable $\chi^2$.
Figure~3 shows one of the linear fits.
 The values of $R$ with statistical
errors at fixed $x$ vs $Q^2$ are shown in Figure~4.
The new data, which extends a decade lower in $x$, are in agreement
with the QCD based model. 
The values of $R$ are not sensitive to the normalization 
of the absolute flux.
The dominant uncertainty at very small $x$ 
and low $Q^2$ is
from the non-zero value of $\Delta xF_3$
because of uncertainties in the
slow rescaling formalism and the level of the strange sea.
Therefore, the values of $R$ 
in the kinematic region $x<0.1$ and $Q^2<4$ GeV$^2$ are not shown
at present.

\begin{figure}
\centerline{\psfig{figure=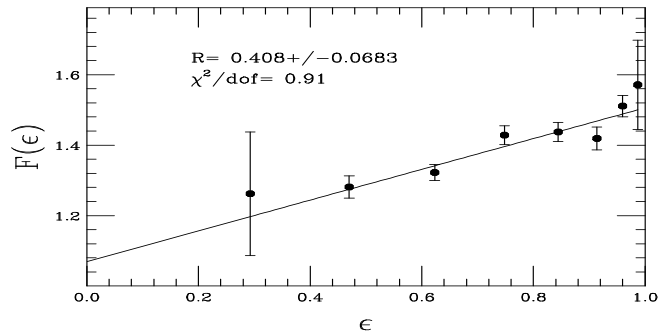,width=3.4in,height=1.7in}}
\caption{F(x,$Q^2$,$\epsilon$) vs $\epsilon$ fit at $x=0.045$ 
and $Q^2=5.01$ GeV$^2$: The errors on $F$ are only statistical errors.}
\end{figure}

Better modelling of $\Delta xF_3$ at low $Q^2$ and small $x$,
and more precise data from the 1996-97 run of NuTeV (FNAL E815)
experiment with 
sign selected neutrino beams are expected to yield better determinations of
$R$ at small $x$.

\begin{figure}
\centerline{\psfig{figure=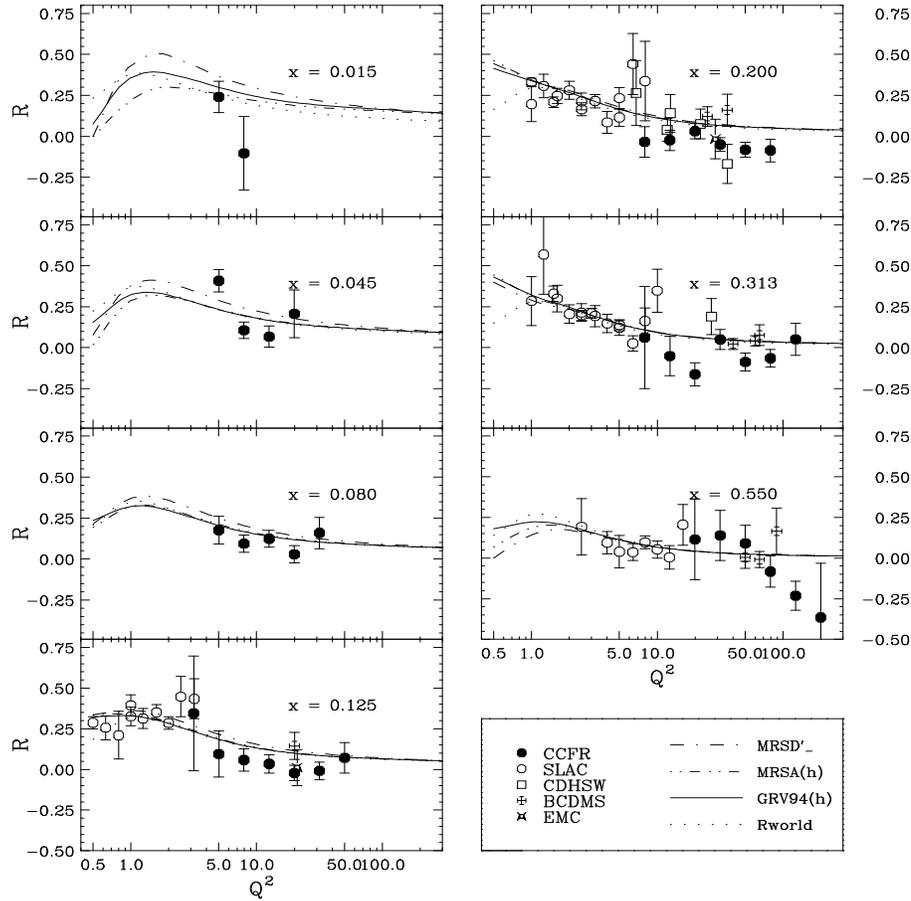,width=4.7in,height=4.7in}}
\caption{$R$ at fixed $x$ vs $Q^2$: Preliminary CCFR data are compared
with other data and
 with the Bodek, Rock and Yang QCD based fit with various parton
distributions.}
\end{figure}

\end{document}